# Revisiting Thin Silicon for Photovoltaics: A Technoeconomic Perspective

Zhe Liu[1,*], Sarah E. Sofia[1], Hannu S. Laine[1,†], Michael Woodhouse[2], Sarah Wieghold[1,§], Ian Marius Peters[1] and Tonio Buonassisi[1,*]

[1]Department of Mechanical Engineering, Massachusetts Institute of Technology (MIT), Cambridge MA USA
[2]Strategic Energy Analysis Center, National Renewable Energy Laboratory (NREL), Golden CO USA
[*]Corresponding authors: Z. Liu (zheliu@mit.edu) and T. Buonassisi (buonassisi@mit.edu)
[†]Now at Sellforte Inc., Helsinki, Finland
[§]Now at Florida State University, Tallahassee, FL USA

## Abstract

Crystalline silicon comprises 90% of the global photovoltaics (PV) market and has sustained a nearly 30% cumulative annual growth rate, yet comprises less than 2% of electricity capacity. To sustain this growth trajectory, continued cost and capital expenditure (capex) reductions are needed. Thinning the silicon wafer well below the industry-standard 160 μm, in principle reduces both manufacturing cost and capex, and accelerates economically-sustainable expansion of PV manufacturing. In this Analysis piece, we explore two questions surrounding adoption of thin silicon wafers: (a) what are the market benefits of thin wafers? (b) what are the technological challenges to adopt thin wafers? In this Analysis, we re-evaluate the benefits and challenges of thin Si for current and future PV modules using a comprehensive technoeconomic framework that couples device simulation, bottom-up cost modeling, and a sustainable cash-flow growth model. When adopting an advanced technology concept that features sufficiently good surface passivation, the same high efficiencies are achievable for both 50-μm wafers and 160-μm ones. We then quantify the economic benefits for thin Si wafers in terms of poly-Si-to-module manufacturing capex, module cost, and levelized cost of electricity (LCOE) for utility PV systems. Particularly, LCOE favors thinner wafers for all investigated device architectures, and can potentially be reduced by more than 5% from the value of 160-μm wafers. With further improvements in module efficiency, an advanced device concept with 50-μm wafers could potentially reduce manufacturing capex by 48%, module cost by 28%, and LCOE by 24%. Furthermore, we apply a sustainable growth model to investigate PV deployment scenarios in 2030. It is found that the state-of-the-art industry concept could not achieve the climate targets even with very aggressive financial scenarios, therefore the capex reduction benefit of thin wafers is advantageous to facilitate faster PV adoption. Lastly, we discuss the remaining technological challenges and areas for innovation to enable high-yield manufacturing of high-efficiency PV modules with thin Si wafers.

## Broader Context

Climate change is among the greatest challenges facing humankind today. Given the urgency of transitioning to a carbon-neutral energy system, we need to accelerate the deployment of existing renewable technology in the near term. With rapid technological progress and cost decline, silicon photovoltaics (PV) modules is a proven technology to be deployed to a multi-terawatt scale by 2030. Despite the high growth rate in the past decade, the capital-intense nature of silicon PV manufacturing hinders the sustainable growth of the industry. Today, the most significant contribution to capital expenditure (capex) of PV module fabrication still comes from silicon wafer itself. Reducing wafer thickness would have a proportionate effect on wafer and poly capex; however, wafer thickness reduction has been much slower than anticipated. This study revisits the concept of wafer thinning in the context of current technology status and cost structure of PV module manufacturing. The state-of-the-art technoeconomic framework is presented to analyze potential economic benefits in terms of reductions in manufacturing capex, module cost and levelized cost of electricity. The sustainable growth model is further adapted to evaluate the impact of thin wafers on potential acceleration of PV deployment. The critical aspects for industrial adoption of thin silicon wafers are discussed.

# 1 Introduction

Thin silicon wafers for photovoltaics have historically attracted attention, especially in the mid-2000s when the shortage of polysilicon feedstock supply caused large price increases [1], [2]. Utilizing less silicon per wafer was recognized as a promising path to reducing capital expenditure (capex) and module cost [3]. However, thin Si wafers failed to gain significant market traction because of collapsing polysilicon prices, low drop-in manufacturing yield of thin Si wafers, and lack of widespread adoption of high-efficiency architectures for thinner wafers. This work offers a fresh look at thin silicon wafers, and revisits the value propositions and challenges with modern solar cell architectures and cost structures.

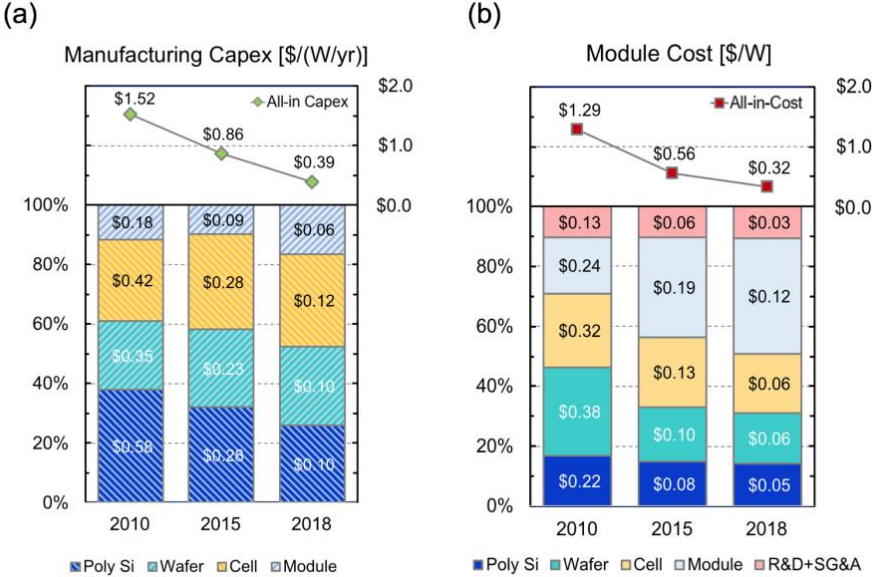

**Fig. 1:** (a) PV manufacturing (*i.e.*, poly-Si to module) capex and (b) module cost for monocrystalline Si PV modules in 2010, 2015 and 2018. Specific contributions in the supply chain of PV manufacturing are broken down into: poly Si production, ingot growth and wafering, cell processing and module assembly. Capex is reported as the capital expense normalized by the annual manufacturing capacity in Watts with the unit of [$/(W/year)], while module cost is reported as production expenses normalized to the module power output in the unit of [$/W]. These values are the NREL benchmarks from Ref. [4] that represent the median of the global manufacturing in the respective year.

Benefiting from efficiency advances, throughput improvements, materials savings and economies of scale, excellent progress has been made in capex and cost reductions (which can be seen in Figure 1 for the data adopted from [4]). From 2010 to 2018, the total capex for PV production from poly-Si to module, which is defined as the total capital normalized by the annual capacity in Watt, has declined by 75% from 1.52 to 0.39 $/(W/year) [4]. Over the same period, the total processing cost from poly-Si processing to module assembly was also reduced by 75% from 1.29 to 0.32 $/W [4]. This cost reduction partially came from efficiency improvements, because the benchmark efficiency of industrial modules has increased from 14% in 2010 to 17% in 2015 and 19% in 2018 in Ref. [4]. The benchmark of 19% module efficiency in 2018 was achieved by widely adopting the technology of Passivated Emitter and Rear Cell (PERC). PERC PV modules are now fabricated more cheaply than conventional Aluminum Back-Surface-Field (Al-BSF) cells, and have become the new industry standard. Besides the impact of efficiency improvements, significant reductions are observed in terms of per-module-area capex and cost, which are shown in Figure S1 of the Electronic Supplementary Information (ESI). We find that the per-area capex declined 65%, from 213 $/(m$_2$/year) in 2010 to 74 $/(m$_2$/year) in 2018; whereas, the per-area cost fell 66%, from 180 $/m$_2$ in 2010 to 61 $/m$_2$.

These achievements are noteworthy but are insufficient to enable the PV industry to meet climate targets defined by the Intergovernmental Panel for Climate Change (IPCC) through PV deployment [5], [6]. Needleman *et al.* [7] estimated that a cumulative PV installed capacity of 7 – 10 TW by 2030 would be required to have a reasonable chance of sufficiently reducing electricity-related carbon emissions and keeping the global temperature rise below 1.5 – 2°C. However, the current PERC baseline is not able to achieve this level of installation by 2030 (see Figure S6 in ESI). Further reduction in capex is needed to sustain the high growth rate of PV installations.

Wafer thickness reduction offers a pathway to effective reductions in both capex and cost, because capex and cost of all manufacturing steps upstream of wire sawing are reduced proportionally with the grams of silicon used per Watt. As seen in Figure 1, the combined capex contribution of the poly-Si and wafering processes have persistently been above 50% over the past eight years. Similarly, the combined cost contribution of the two processes has been reduced, but still accounts for over 30%. From this perspective, reducing wafer thickness appears promising to reduce capex and cost. There are two key questions still to be addressed: (1) How much can we still benefit economically today from the "old" idea of wafer thickness reduction? (2) What are the technologies needed to produce high-efficiency thin Si modules with high production yield and high power-conversion efficiency?

To answer these two questions, we apply technoeconomic modeling to quantify the potential cost and capex benefits of thin silicon manufacturing, and surveying technology pathways that enable manufacturing with high yields and efficiencies. We revisit the efficiency vs. thickness trade-off in the light of recent advances in cell architecture, which should, in theory, push the critical thickness for maximum efficiency to lower values. We also quantify economic benefits and the ability to meet climate targets if the industry successfully adopts thin wafers. Lastly, we analyze remaining technological barriers for thin wafers, especially those pertaining to manufacturing yield.

## 2 PV device simulation: Effect of wafer thickness on efficiency

Efficiency is an impactful factor for both capex and cost reductions [8], [9]. Thinning wafers reduces their ability to capture available photons, especially those in the near-infrared spectral range. As a result, short-circuit current may be reduced, and therefore there is a concern of an efficiency penalty. However, it was noticed that efficiency loss due to lower short-circuit current can possibly be compensated by the increase of open-circuit voltage and fill factor if the surface passivation is sufficiently good. Many of these previous studies [10], [11] about efficiency versus wafer thickness were conducted when the mainstream industry devices were Al-BSF cells, which have poor rear surface passivation. The recent industrial transition to PERC aims to reduce rear surface recombination. With further advancement of surface passivation technology, recent studies [12], [13] suggest that wafer thinning may no longer be as detrimental for conversion efficiency. Better passivation in the rear surface also coincides with an improved optical performance, which also contributes to higher efficiency. To quantify the relation of efficiency versus wafer thickness, we performed a set of comprehensive but generalized device simulations in PC1D [14]. A total of four device concepts are considered in our simulations, namely Al-BSF, PERC, advanced PERC+, and advanced high-efficiency technology (HE-Tech). Many of these advanced device concepts have two- or three-dimensional architectures. We used device models with effective simulation parameters [15] in order to resemble the performance of these advanced concepts. Figure 2 shows the simulated module efficiencies depending on the Si wafer thickness.

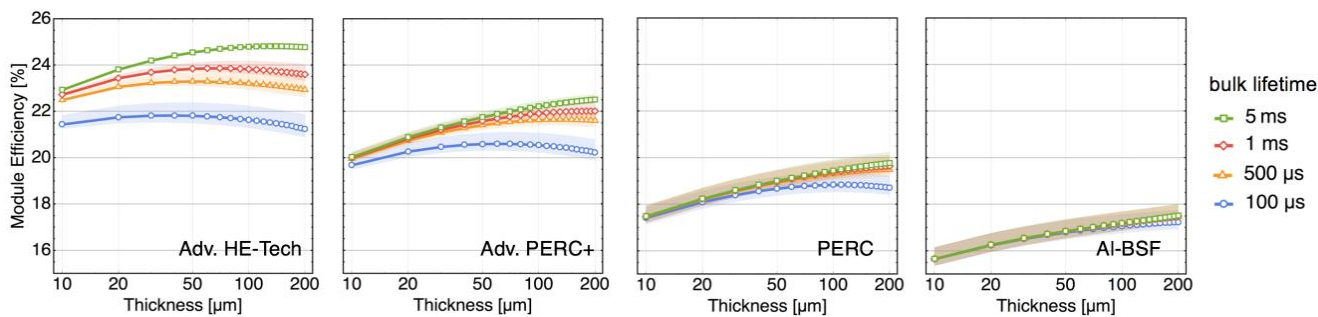

**Fig. 2:** Simulated module efficiency versus Si wafer thickness (*p*-type) for four solar cell concepts (*i.e.*, advanced HE-Tech, advanced PERC+, PERC and Al-BSF). See Table I for the simulation parameters. Lines with symbols mark efficiencies of 2 Ω·cm wafers, and shaded areas indicate efficiencies for a wafer resistivity range of 1 – 3 Ω·cm. In the example of 1-ms bulk lifetime, the relative efficiency ratios of 50 μm to 160 μm thicknesses are 101%, 98%, 97%, and 97% for advanced HE-Tech, advanced PERC+, PERC and Al-BSF respectively.

For simulations of all four device concepts, the bulk lifetime of the wafer is varied: 100 μs, 500 μs, 1 ms, and 5 ms. For state-of-art *p*-type high-performance multicrystalline Si wafers, bulk lifetimes of 250–500 μs are usually found. In comparison, monocrystalline Si wafers usually have lifetimes that are between 1 ms and 5 ms for high-efficiency concepts. In our analysis, because of the similar trends of *p*- and *n*-type Si (see Figure S2 in ESI for *n*-type simulation results), we focus on the results for *p*-type Si for the ease of comparison with historical data. In addition, from conventional Al-BSF to advanced HE-Tech, the cell-to-module (CTM) efficiency factor is also gradually increased (from 0.83 to 0.92) to reflect improvements in module technology, *e.g.*, light scattering ribbons and backsheets, and multi-wire interconnections. The following descriptions summarize the characteristics of each cell concepts, including the key differences in simulation parameters. More detailed parameters can be found in Table S1 in ESI.

a. Conventional Al-BSF solar cells have high effective rear surface recombination velocities (SRV) of around 1000 cm/s, and low rear internal reflectance of around 65% [16]. The poor surface passivation and high parasitic absorption become the efficiency limiting factor in these cells. In fact, PV industry has largely moved to PERC because of higher efficiency. Al-BSF architecture is still considered as a historical reference to demonstrate the decreasing efficiency trend with lower wafer thickness.

b. Current industrial PERC solar cells feature rear passivation through an $AlO_x/SiN_x$ dielectric stack, which has an effective rear SRV of around 100 cm/s (with a typical range of 50 – 200 cm/s) [16]. Light management is also improved by optimizing layer thicknesses of the $AlO_x/SiN_x$ stack, and therefore excellent rear internal reflectance is achieved. The value of 93% is used in all the following device concepts. Despite some improvement in rear surface passivation, the loss analysis studies [17], [18] still suggest that recombination at the rear surface (especially at the rear metal-Si interface) is the efficiency limiting factor of the PERC architecture.

c. One next-generation device architecture, here called "advanced PERC+", marks an advancement of the current PERC structure via further rear passivation improvement. The rear passivation in advanced PERC+ is shown to be another order magnitude lower in SRV than PERC, reaching around 10 cm/s (with a typical range of 5 – 30 cm/s [19], [20]). This reduction in rear SRV could be achieved by eliminating the recombination at metal-Si interface of the local rear contacts in PERC. Therefore, device architectures for advanced PERC+ concepts are likely to be contact-passivated solar cells, *e.g.*, tunnel oxide passivated contacts (TOPCon) [18], poly-Si on oxide (POLO) contacts [21], fired passivated contacts (FPC) [22] and heterojunction with intrinsic thin layer [23]. Recently, both Jinko Solar and Trina Solar launched their TOPCon cells and modules (the best cell efficiency >24% leading to the expected module efficiency around 21%) [24], [25], and REC Solar also launched their heterojunction module (with the best module efficiency of 21.7%) [26]. These technologies utilizing contact-passivated solar cells have the potential to bring module efficiency above 22%. Therefore, the efficiency limiting factors in the advanced PERC+ start to shift to Auger recombination at highly doped region, optical shading of front metal contacts, and front surface recombination.

d. The advanced HE-Tech architecture represents a concept that surpasses the advanced PERC+. These device concepts further reduce the surface recombination on both surfaces, as well as Auger recombination in the highly doped emitter regions. Our simulation assumed values that are an order of magnitude below those of the advanced PERC+ for emitter doping concentration and SRVs (both front and rear). This simulation model is in fact an effective model which captures the key features in a simplified architecture. In practice, HE-Tech concept requires more complex architectures, such as, 26.1% *p*-type IBC solar cell with POLO [27], 25.7% *n*-type both-side-contacted solar cell with TOPCon [28] and 26.3 – 26.7% *n*-type IBC solar cells with include Si heterojunction HIT architecture [29]–[31]. Industrial-size large-area module with IBC cells have achieved >24% record efficiency, for example, 24.1% achieved by SunPower in 2016 [32] and 24.4% achieved by Kaneka in 2017 [31]. However, commercialization of these module technology requires further R&D efforts on cost reduction.

From device simulation results shown in Figure 2, we find that significant efficiency losses with thinner wafers are a concern for conventional Al-BSF or current PERC, but the efficiency loss is less evident for some advanced concepts with better surface passivation. For example, the optimum efficiency for the advanced HE-Tech concept is around 50 μm for bulk lifetimes of 500 μs and 1 ms; whereas the optimum efficiency for advanced PERC+ is found around 100 μm thickness for bulk lifetimes of 500 μs and 1 ms. For the highest bulk lifetime of 5 ms in advanced HE-Tech devices, no significant efficiency reduction appears above 100 μm thickness.

## 3 Potential economic benefits of thin silicon

### 3.1 Cost modeling of thin silicon wafers

Device simulations in the previous section showed that it is possible to preserve efficiency while moving to thinner wafers. With advanced device concepts, we see a clear benefit of moving to thinner wafers because of the reduced material usage without significantly sacrificing performance. After obtaining the efficiency versus thickness relations, we can attempt to answer the first question of this work posted earlier: how much economic benefit can we still obtain by thinning wafers? Herein, we quantify benefits in capex, module cost and levelized cost of electricity (LCOE).

To do so, we utilized bottom-up capex and cost models developed by Powell *et al.* [8], [9] and updated the models with the cost numbers from the recent NREL PV cost analysis report [4]. These cost numbers represent the estimated median of global PV module productions in H1-2018. The benchmark case corresponds to a PERC module with 160 μm monocrystalline Si wafers and 19% efficiency. In our current cost model, efficiencies were varied according to the previous simulation results in Figure 2, and a fixed rate of Si utilization (~63%) was assumed for all varying thicknesses. The current 160 μm thick mono-Si wafer has a kerf loss of 95 μm. The historical trend shows the kerf loss by wire sawing process has been steadily reduced (see Figure S9 in ESI), with a projection of further reduction to its technology limit. A 100-μm-thick wafer will correspondingly have a kerf loss of 60 μm, which is approximately the predicted technological limit of diamond wire sawing in the ITRPV report [33]. Achieving thinner wafer with a kerf loss less than 60 μm will require a new alternative wafering technology (*e.g.*, kerfless wafer growth). Because our focus is mainly on the potential impacts of thickness and efficiency variations, we assumed other variables (*i.e.*, fixed costs of properties, plant and equipment (PPE) and variable costs of materials and processes) are kept fixed at the benchmark values in 2018. This assumed scenario provides the analysis of cost reduction potential, echoing a previous NREL study [3]. Some other cost scenarios, such as, constant kerf loss and, increased capex and cost, are analyzed in ESI. In addition, uncertainties of all variables will be discussed in the next subsection (Section 3.2), and available technologies and present challenges will be discussed in detail in Section 4.

Furthermore, we also conducted LCOE analyses for utility-scale PV electricity systems in the United States. We used baseline values of energy yield and balance-of-system costs from Ref. [34], [35], which correspond to the median LCOE scenario in the United States in 2018. The module prices in this LCOE scenario are updated by the simulated module costs plus a 15% operating margin [4]. The modeling results of module capex, cost and LCOE versus thickness are shown in Figure 3a, 3b and 3c respectively. The cost analysis of conventional Al-BSF is not considered here because mainstream of PV industry has transitioned to PERC devices. Note that all capex, module cost and LCOE models for this work can be found in Excel spreadsheets in the ESI.

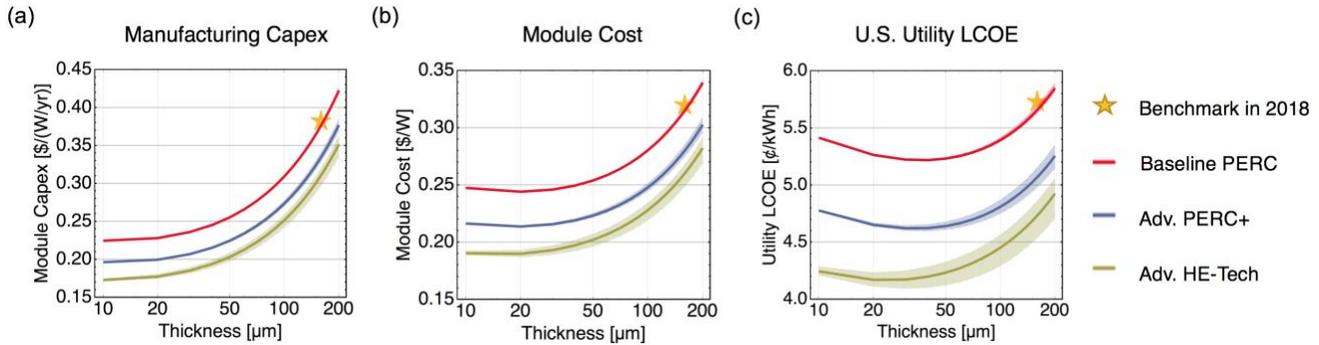

**Fig. 3:** Module capex, cost and US utility LCOE analysis of the three solar cell concepts (PERC, advanced PERC+ and advanced HE-Tech) at different Si wafer thicknesses. The solid lines represent the baseline cases of the 1 ms bulk lifetime, and the shaded areas indicate the cost variations for the bulk lifetime range of 500 μs – 5 ms, and. The capex and cost analyses are based on the median costs of global production from Ref. [4]. The utility LCOE analysis is based on the median cost structure of utility PV system in the United States [34]. Benchmark values are for 160-μm-wafer PERC modules in 2018.

From Figure 3a and 3b, we see the as-expected reductions in capex and cost via the savings of silicon material, as well as via efficiency improvements. Both capex and cost monotonically decrease with wafer thickness down to 20-μm (or even less) regardless of the technology concept. For example, reducing wafer thickness from 160 μm down to 50 μm for the current PERC concept can potentially get a capex reduction of ~0.14 $/(W/year) and a cost reduction of ~0.07 $/W. This means, for every 10-μm thickness reduction, manufacturing capex declines roughly by 1.3 ¢/(W/year) and module cost declines roughly by 0.6 ¢/W. According to the LCOE analysis in Figure 3c, for all three device architectures, utility LCOE minima are located at the wafer thickness of ~50 μm (with a range of ±20 μm). We find that thin silicon can reduce LCOE by more than 5% relatively from the value of 160-μm wafers, regardless of the device technologies. The 5% reduction in LCOE is in fact very substantial for the industry to make a change. To put it in context, PV industry has transitioned from Al-BSF to PERC to harness the 3% reduction in LCOE [36].

To better understand these results, it is important to know the different impacts these cost factors have on the PV industry. Module capex largely affects the growth rate of PV manufacturing industry, and therefore lower capex industries tend to have higher self-sustained growth rates. Module cost typically affects the competitiveness of a certain type of PV module. In the past, it has been very difficult to gain market traction with modules that have improved efficiency at a higher cost. LCOE, which is influenced by both module cost and efficiency, affects the competitiveness of PV electricity at a specific location. Therefore, lower LCOE incentivizes consumers to adopt more PV systems. We observe that the recent technology transition from Al-BSF to PERC ultimately started when PERC became cheaper in all three cost factors (in addition to featuring higher efficiencies). Furthermore, we acknowledge that our models are rather simplified for future advanced technologies. In fact, it is very difficult to build a bottom-up cost model accurately without clarity about what technologies will be used for thin Si. To account for the variability of the assumed parameters, we conduct an uncertainty analysis of the cost and capex models in the next sub-section.

## 3.2 Uncertainty analysis of the cost and capex models

The cost analysis in this study focuses on the maximum potential impacts of efficiency changes and of the amount of silicon usage. Other factors are assumed to remain constant (*i.e.*, do not contribute to capex and cost reduction). The only exceptions are the cost of Selling, General and Administrative (SG&A) and R&D, which were usually considered as a constant percentage of the total cost. Therefore, the six parameters that are likely to change in the cost model are efficiency improvement, the amount of silicon saving, SG&A and R&D, manufacturing yield, variable cost, and direct PPE expense. We conducted an sensitivity analysis on a target scenario: the advanced HE-Tech with 50-μm wafer thickness. This is the scenario where the optimum efficiency of 23.8% is achieved (red curve in Figure 2a). At the same time, its LCOE (~4.2 ¢/kWh) is also close to the minimum (green curve in Figure 3c). Table I shows the target scenario of the advanced HE-Tech module with a 50-μm wafer and the current PERC module with a 160-μm wafer.

Table I. Two selected scenarios of the silicon solar modules

| **Module parameters** | **Baseline PERC** | **Advanced HE-Tech** |
|---|---|---|
| Module efficiency | 19.0% | 23.8% |
| Thickness | 160 μm | 50 μm |
| Kerf loss | 95 μm | 28 μm |
| Si usage per Watt | 3.1 g/W | 0.77 g/W |
| Manufacturing capex | $0.39 /(W/yr) | $0.20 /(W/yr) |
| Module cost | $0.32 /W | $0.20 /W |
| U.S. utility LCOE | ¢5.5 /kWh | ¢4.2 /kWh |

The results of the uncertainty analysis are shown as two tornado charts in Figure 4. The changes in capex and cost were obtained in response to the ±5% relative change in each specific factor. Uncertainties are ranked from high to low. The accumulated uncertainty on the predicted capex and cost of the advanced HE-Tech concept is also shown as the lowest bar in Figure 4. It shows that, if all six parameters vary simultaneously by 5%, both capex and cost will have a combined range of uncertainty up to ±20% from the calculated values in Table I. Both, capex and cost, are very sensitive to manufacturing yield and efficiency with a nearly 1-to-1 sensitivity. This agrees with the previous findings by Powell *et al.* [37]. Si usage affects the capex and cost to different extent because of its different proportion to the total capex and cost. PPE variation results in a 1-to-1 change in total capex and a smaller change in module cost via the depreciation of PPE. The variable cost uncertainty only changes the module cost (and has no impact on capex). Costs of SG&A and R&D are a part of the operating cost, which only influences module cost. From the ranking in Figure 4, we identify the most critical aspects to achieve this target scenario of advanced HE-Tech with 50 μm wafer: (1) Realize efficiency improvement in mass production; (2) Maintain high production yield; (3) Maintain low manufacturing cost and low equipment capex of the production; (4) Realize the amount of the assumed Si saving. The challenges and these factors will be discussed in full detail in Section 5.

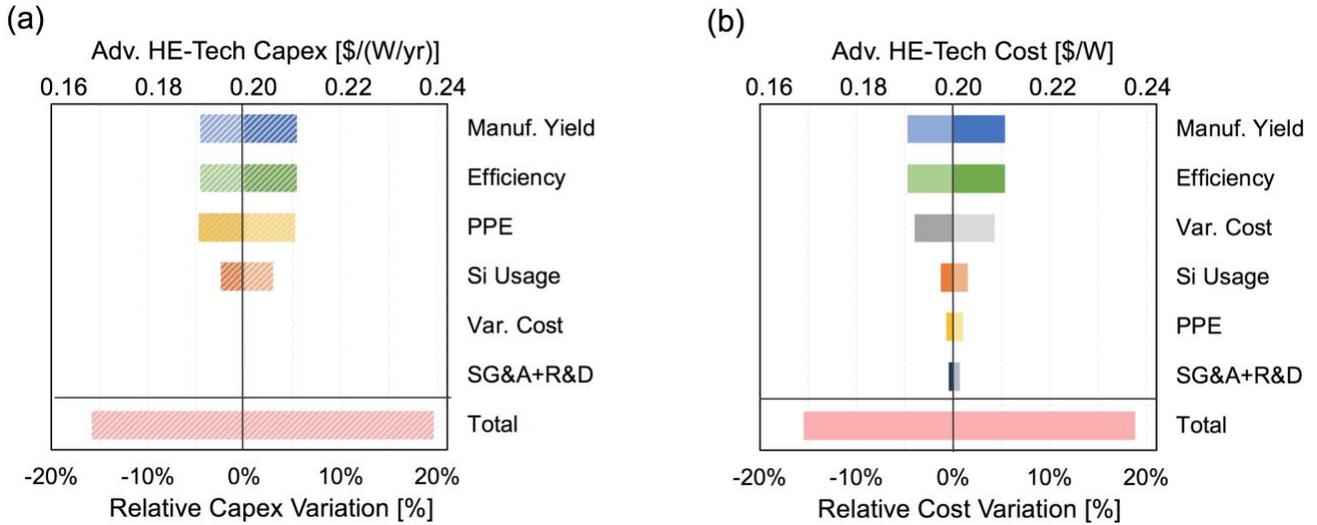

Figure 4: Uncertainty analysis for an advanced HE-Tech module with 23.8% efficiency using 50-μm-thickness wafers. The six parameters under investigation are efficiency improvement, silicon saving, cost of SG&A and R&D, manufacturing yield, variable cost and PPE. The bars indicate capex or cost changes in response to a ±5% change of every specific factor. Lighter color indicates an increase in the parameter, and dark color indicates a decrease in the parameter. The "Total" range of uncertainty indicates the combined effect on capex or cost if all six parameters are changed by ±5% simultaneously and the positive and negative responses (in capex and cost) are summed up separately.

## *3.3* Implications for sustainable PV growth

To shine some light on the practical implications of these capex and cost values for the growth of PV deployment, we used the demand-constrained growth model that was described previously in Ref. [7]. It was previously recognized that, capex reduction is very important to sustainable growth of PV industry [9]. With the potential capex and cost reduction in the example of advanced HE-Tech module with 50-μm-wafer and 160-μm-wafer, we want to investigate the potential boost in PV growth in comparison to our current PERC technology with 160-μm wafers. For this purpose, we simulated PV growth scenarios for these two cases with the goal of achieving the IPCC 2030 climate targets.

The basic idea of this growth model is that PV growth is constrained by two factors: installation demand and production capacity (*i.e.*, PV module supply). Whichever of those two factors is lower limits growth. Installation demand is set by an empirical function of module price [38], whereas the rate of adding new production capacity depends on the ratio of cash (*i.e.*, profit plus debt) to capex. With this model, we assessed PV growth for several scenarios of operating margins [*i.e.*, the percentage earning before interests and taxes (EBIT)], and debt-to-equity (D/E) ratios (*i.e.*, the ratio of debt borrowing over cash earning). Typically, the operating margin in the PV industry is very volatile and determined as a result of market conditions. Many PV companies are found to have had operating margins below 10% in the past five years [3]. Only a few companies, such as GCL-Poly and Longi, were able to maintain operating margins above 20% in recent years (see Figure S7 in ESI). However, there is a need to point out that a good operating margin is important to attract new investments and sustain new additions of production capacity [39]. On the other hand, D/E ratio can, to some extent, be decided by PV companies based on their expansion plans. Due to severe competition and rapid market expansion, most PV companies leverage higher debt in the capacity expansion (with D/E ratios of up to 5), which can be seen in their annual cash flow statements (according to the data from SEC company filings [40]). High debt for expansion was also reported previously by Chung *et al.* [41]. It is generally very common that companies in the growth phase of their life cycle tend to have lower operating margins due to competition and leverage higher debts for market expansion [42].

In this work, we varied operating margins and D/E ratio in a relatively wide range to evaluate possible scenarios to achieve more than 7 TW cumulative PV installation in 2030. For operating margin, a value of 15% was used as a baseline in previous cost analyses [4], [7]. The values of 10% and 20% represent lower and higher margin scenarios. The very low margin of 5% marks a case where growth is nearly impossible. In previous studies [7], [9], a D/E ratio of 1 was assumed as a low-risk reference. Due to high capital intensity of the PV industry, we also consider D/E ratios of 2 and 5 as intermediate and aggressive leverage of debt scenarios. Simulated cumulative installation by 2030 is shown in Figure 5 for different values of operating margin and debt ratio. Note that the simulated results using this growth model approximate the upper limit in each scenario by assuming immediate adoption of the assessed technology, and reinvestment of all the profits into capacity expansion.

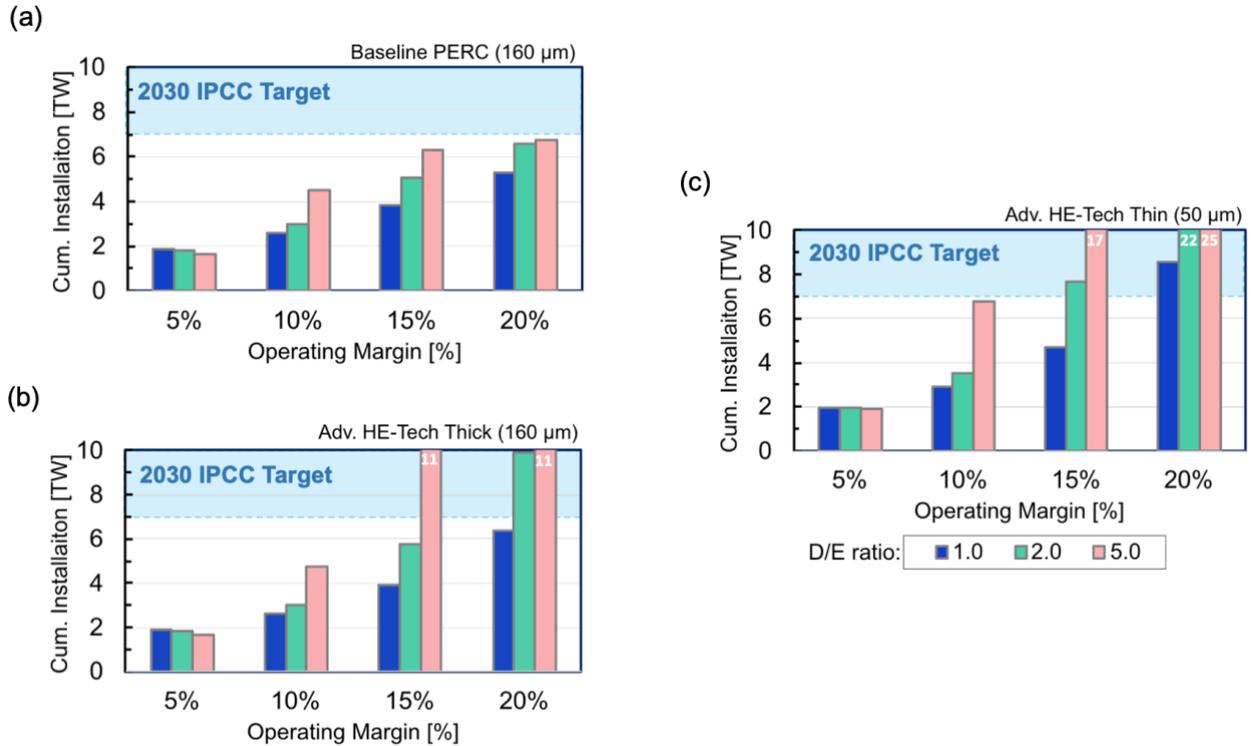

**Fig. 5:** Simulated cumulative PV installation by 2030 of three investigated solar cell concepts (*i.e.*, baseline PERC with 160 μm wafer, advanced HE-Tech with 160 μm wafer, and advanced HE-Tech with 50 μm wafer) using a demand-constrained growth analysis. Four different operating margins were used: 5%, 10%, 15% and 20% of the module selling price, and three different debt-to-equity (D/E) ratios for new capacities: 1:1, 2:1 and 5:1. The light blue shaded area indicates the 7–10 TW peak PV capacity that was determined in [7] as a climate target for 2030, following suggestions from [7] based on IPCC climate targets [4].

In Figure 5, we observe that the baseline PERC architecture cannot achieve a cumulative installation of more than 7 TW for all considered scenarios. At 20% operating margin and 5x debt, it can only get close to 7 TW. At the same 5x debt but only 10% operating margin, the advanced HE-Tech with 50 μm wafer can achieve a similar amount of cumulative installation. However, growth with such a high debt ratio is not sustainable in the long term. Increasing debt can increase growth very efficiently in the short run, but aggressive debt leverage significantly increases the company's financial risk with current volatile module prices. Based on current margins, it is not very likely that these ambitious growth targets are achieved. Realizing the climate goal by 2030 with additional debt is, therefore, not desirable as it may prevent the industry from keeping up with the long-term growth in electricity demand. In Figure 5b of HE-Tech with 160 μm wafer, we observe that efficiency improvement could possibly bring

PV growth to 10 TW with 20% operating margin and moderate debt of 2x. In comparison, HE-Tech with 50 µm in Figure 5c could reach 22 TW under the same condition of 20% operating margin and moderate 2x debt. Other more sustainable growth scenarios for the advanced HE-tech with 50 µm wafer are 20% operating margin with only 1x debt to achieve 8.6 TW in 2030 or 15% operating margin with 2x debt to achieve 7.8 TW in 2030. We acknowledge that the current operating margin is very low (less than 10%), but it is the hope that a premium (with higher operating margin) can be charged for PV modules when LCOE is much lower than for other forms of electricity generation (*e.g.*, natural gas plants). Furthermore, because of the significant cost and capex reduction with thin wafers, HE-Tech with 50 µm wafer has better long-term growth potential than the one with 160 µm wafer (see Figure S6 in ESI).

# 4 Why is thin Si not here yet? Challenges & innovation opportunities

Our analysis elucidates that we could achieve lower cost, lower capex, and high-efficiency next-generation PV modules with thin silicon wafers. What are the technology developments needed to achieve the ultimate paradigm shift toward thin wafers? Figure 6 summarizes some important key areas across the manufacturing supply chain. These technology areas echo the four aspects that are identified previously from the cost analysis in Section 3.2.

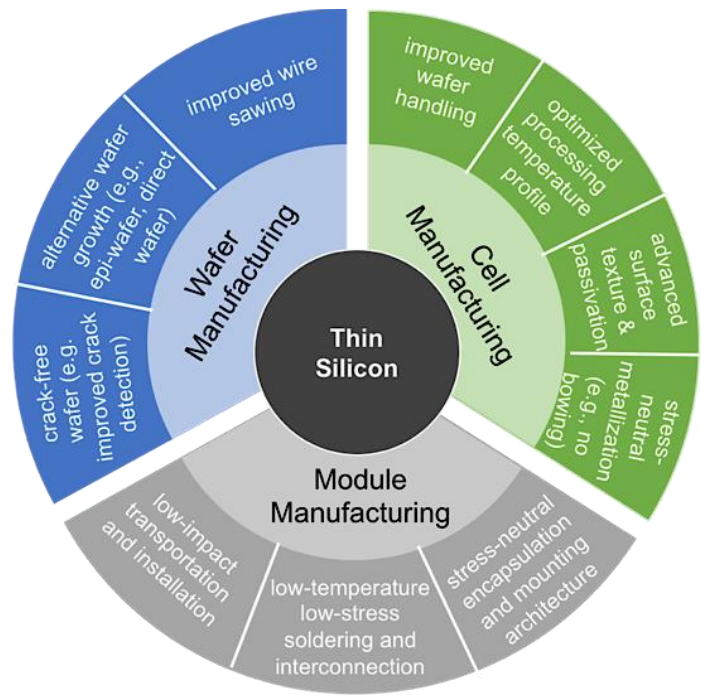

**Fig. 6:** R&D areas to achieve the high-efficiency thin-wafer PV modules (with low cost and capex).

a. **Concern of production yield loss.** Fabricating PV modules with thin wafers is very challenging. High wafer breakage rates are found at various stages of manufacturing, installation, and field operation. The main cause of yield loss during manufacturing comes from the stress and strain induced when handling wafers, cells and modules. Many tool adjustments are needed to handle thin wafers in today's manufacturing line. After fully optimizing the process steps, Harrison *et al.* [43] demonstrated a reasonably good yield of ~93% down to ~90-µm wafers on HIT production line, but they still faced less than 80% yield for the thickness below 80 µm. However, we noted that many handling steps in the improved procedures in [43] were completed manually, which will scale up to mass production. Thus, the first key innovation area for yield loss reduction is to improve wafer handling technologies, such as non-contact Bernoulli gripper [44]. Wafer handling is even more essential for wafers thinner than 80 μm due to the extremely high breakage rate. Completely new fabrication processes may be needed. One possible method is to manufacture solar cells with kerfless wafers on a supporting carrier [45]–[47]. In addition, the presence of microcracks is known to reduce the wafer strength [48], and the critical crack size becomes much smaller for thinner wafers [49]. Improved microcrack inspection could help identify the process steps where microcracks initiate [50]. During module installation and field operation, one of the key root causes of breakage is the local stress induced by interconnection wire. Multi-wire interconnection, which is shown to induce lower stress on solar cells, is suitable for thin wafers. Furthermore, instead of the conventional front-to-back zig-zag connection, researchers have demonstrated some examples of new interconnection schemes for stress reliving [51], [52]. All in all, given the criticalness for transiting toward thin wafers, we would like to emphasize the urgency and importance of finding manufacturing solutions. In fact, innovative researches on new technologies to avoid yield loss for thin wafers have not been paid sufficient attention to in the past decade.

b. **Concern of efficiency penalty**. Efficiency penalty comes from the incomplete absorption of NIR photons when the thickness is reduced. It is because Si is an indirect semiconductor and requires a relatively long optical path to absorb near-bandgap photons. However, achieving high efficiency with advanced surface passivation [53]–[55] should not be the limiting factor for thin silicon, as indicated by our device simulations. Previous successes of >20%-efficient solar cells have been demonstrated for less-than-100-μm-thick wafers. For example, solar cell efficiency of >20% was achieved with diffused-junction technology (*i.e.*, PERC or its derivatives) with using 80 μm wafers in large batches [56], [57]. Sanyo produced a 24.6% Si HIT cell on 98 μm with industrially compatible tools [23]. For ultra-thin wafer around 50 μm thickness, many feasible concepts [58], [59] were also demonstrated with small-area solar cells (~4 $cm_2$). One notable result is that Solexel achieved an efficiency of 21.2% for full-size solar cells with 35 μm thick kerfless wafer [60]. Many new architecture concepts of device architecture discussed in Section 2 can be utilized for thin-wafer-based PV modules without the efficiency penalty. With further advancement of new light management schemes, *e.g.*, black silicon with nanoscale textures [61]–[64], the loss due to incomplete absorption in thin wafers can be reduced. In addition, innovations of new encapsulation materials could also enhance NIR light trapping at module level [65]–[67]. In summary, PV R&D is heading to the direction to achieve excellent surface and contact passivation [68], which will reduce or even eliminate the efficiency penalty and benefit the transition to thin wafers.

c. **Concern of additional PPE expense and variable cost**

With higher quality materials and more sophisticated device architectures, advanced technology concepts with thin wafers may necessarily require additional PPE expenses and variable costs for manufacturing. However, learning from historical trends (Figure S1 in ESI), new technologies are required to be produced more cheaply in terms of per-area cost and capex for higher module efficiency in order to gain sufficient market traction. The decreasing trend is, according to ITRPV reports [28], driven by equipment and process innovations to achieve higher throughput, lower material usage, less material waste, and simpler process steps, etc. Therefore, the decreasing trend should not be taken as granted for future technology. Instead, continuous R&D efforts are needed to ensure technology innovations for thin wafers in Figure 6 fulfill these criteria. One example is that the newly developed device architecture with passivated contacts (*e.g.*, TOPCon) may have a slightly better chance to be adopted more quickly than HIT or IBC, because it can better utilize current high-throughput industrial processes (*e.g.*, plasma enhanced chemical vapor deposition) [42], [43]. Another example is that low-stress multi-wire interconnection may require a more sophisticated tabbing and stringing tool, but it offers the advantage of significant savings of silver paste [61], [62]. In conclusion, the PV community has a track record of fabricating better performing solar modules with lower PPE expense and variable cost. To fully extract the economic benefits from thin silicon, a key focus of innovations is on those technologies that maintain low-cost and low-capex manufacturing.

d. **Feasibility concern for thinner wafer production**.

Technologies of making thin wafers down to 100-μm thickness are within the line of sight. For example, Longi Silicon announced slicing 110-μm-thick mono-wafers in their R&D facilities, with the ability to transfer the process to mass production [69]. Terheiden *et al.* [56] demonstrated an industrially-compatible process to making 90 – 100 µm thick mono-wafers via optimizing diamond-wire sawing process. Further thickness reduction to 50 µm or thinner wafers may require novel kerfless wafer growing processes, such as epitaxial mono-wafer (*e.g.*, NexWafe) and directly-grown multi-wafer technologies (*e.g.*, 1366 Technologies). These kerfless wafers have not yet been adopted at a large scale, mostly because of a lack of market for thin wafers. Kerfless wafer manufacturers have to produce and sell wafers with standard 160 – 180 µm thickness, which limits the full advantage of their technology. With the inevitable trend of utilizing thinner and thinner wafers, we may ultimately turn to these viable kerfless technologies to extract the maximum silicon savings possible.

# 5 Conclusion and outlook

In this work, we evaluated the market potential of thin silicon wafers using a technoeconomic framework. First, we compared the efficiency-versus-thickness relations for four device concepts (conventional Al-BSF, state-of-the-art PERC, advanced PERC+, and advanced High Efficiency-Tech) on the module level via numerical device simulations. Secondly, using the simulated efficiency-versus- thickness relations as inputs, we evaluated the potential economic benefits of thinner wafers for state-of-the-art and future advanced technologies. We performed cost modeling analyses of PV manufacturing with the most recent global-median cost numbers of 2018, and observed that cost ($/W) and capex [$/(W/year)] decrease monotonically with wafer thickness. For example, reducing wafer thickness from 160 µm to 50 µm reduces capex by ~0.14 $/(W/year), and cost by ~0.07 $/W for the current PERC module. In comparison, the 5% absolute efficiency increase from current PERC (19%) to advanced HE-Tech (24%) only brings capex down by ~$0.08/(W/year) and cost down by $0.07/W. Thirdly, we also performed an LCOE analysis for the utility-scale PV system in the United States. Efficiency improvements have a strong influence on LCOE because of their implications on BOS costs per Watt. However, we still find significant LCOE benefits even for thickness reduction alone. For all device concepts investigated, the optimal LCOEs occur at wafer thickness of around 50 µm, with only a small variation between device concepts. The LCOE with 50-µm thickness is 5% lower than their 160-µm counterparts, which is slightly larger than the 3% LCOE benefits by transitioning from Al-BSF to PERC [36].

Uncertainty analysis of the six key inputs parameters in the cost models was conducted for a target scenario of HE-Tech with 50 µm wafer thickness. The cost model suggests that HE-Tech modules with 50 µm wafers could potentially achieve a capex of 0.2 $/(W/year) and cost of 0.2 $/W, in comparison with a capex of 0.39 $/(W/year) and cost of 0.32 $/W for PERC modules with 160 µm wafers. In the case of simultaneous ±5% variations of input parameters, we see an uncertainty range up to ±20% for both capex and cost. Furthermore, in order to give a broader perspective on how the cost and capex reductions benefit the PV industry, we performed an industry growth analysis to investigate different scenarios for advanced HE-Tech with 50 µm wafers. Under 15% operating margin and debt ratio of 2, thin wafers can help the PV industry reach close to 8 TW cumulative PV installations by 2030, in comparison with 5 TW for the PERC baseline. Lastly, we evaluated the technology readiness for thin silicon and discussed the challenges of thin silicon around the four most sensitive parameters to affect capex and cost (*i.e.*, module efficiency, manufacturing yield, the feasibility of fabricating thin wafers and low-cost low-capex processing).

Climate change is a pressing challenge, and the PV community has the potential to address it by contributing more carbon-neutral electricity [70]–[72]. Even with today's cost structure, we show that the adoption of thinner wafers still provides very significant capex reductions and considerable cost reduction. Excellent surface passivation and light management are required to minimize the efficiency loss for thin wafers. Given the proliferation of dielectric passivation tools coupled to industrial adoption of passivated cell architectures, the industry is much better positioned to achieve high-efficiency with thin Si wafers today than it was in the mid-2000s. This area is ripe for innovation, a sentiment echoed by forward-looking industry players. Now, the main barrier that hinders the widespread adoption of thin wafers is likely to be manufacturing yield loss. We believe the industry is ready for thin wafers, but an extra effort on developing innovative manufacturing equipment and processes is necessary to overcome this barrier.

# Acknowledgements

The authors thank Dr. Qi Wang from Jinko Solar, Dr. Yifeng Chen from Trina Solar, Dr. Hongbin Fang from Longi Solar and Dr. Shaffiq Jaffer from TOTAL to provide the industrial perspectives about the thin wafer manufacturing. The authors acknowledge the informative discussions with Zekun Ren from Singapore-MIT Alliance for Research and Technology about device simulation, and Dr. Olindo Isabella regarding the recent progress of light management and surface passivation technology. The authors are especially grateful to Dr. Ashley Morishige and Dr. David Berney Needleman, who are former members of MIT PV Lab, for their great effort in initiating this project on thin silicon. This work was partially supported by the U.S. Department of Energy (DOE) under Photovoltaic Research and Development (PVRD) program under Award no. DE-EE0007535. Z. Liu acknowledges partial support from a TOTAL Energy Fellowship through the MIT Energy Initiative. I. M. Peters acknowledges support from the DOE-NSF ERF for Quantum Energy and Sustainable Solar Technologies (QESST) and from Singapore's National Research Foundation through the Singapore MIT Alliance for Research and Technology's "Low energy electronic systems (LEES) IRG".

# Electronic Supplementary Information:

## A. Capex and Cost per Unit Area

The historical data of cost and capex per area PV modules shows the decreasing trend over the years in Figure S1. The per-area cost and capex decouple the effect of module efficiency. The benchmark module efficiencies are 14% in 2010, 17% in 2015, and 19% in 2018. Despite the efficiency improvement, the cost and capex for cell and module processing are kept decreasing. The exponentially fitted trendlines extrapolate the decreasing trend to the future, although the decreasing rate will be slower. This indicates that it is likely to require the new technology to have cheaper processing cost and capex to gain sufficient market traction.

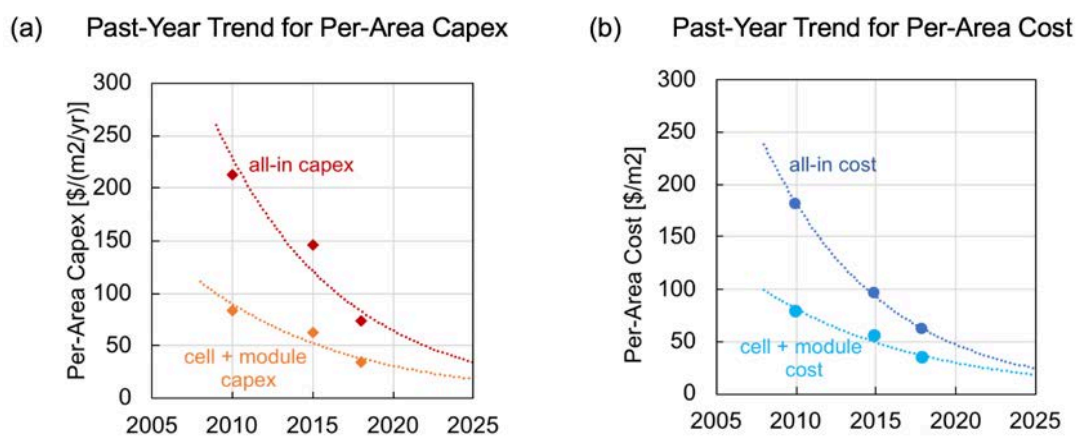

Figure S1: The decreasing trend for per-area capex and cost *versus* years. The capex and cost of the cell and module processes are shown separately in comparison with the all-in capex and cost. The dashed lines are the data trendlines with exponential fitting.

The per-area capex and cost values *versus* the wafer thickness is shown in Figure S2, which directly shows the cost change due to the savings from reducing silicon usage (regardless of the module efficiencies). The cost structure corresponding to the production process of the current PERC modules. We see capex reduction rate is $2.3/(m$^2$/year) for reducing every 10 μm Si thickness, whereas cost reduction rate is $1.2/m$^2$ for reducing every 10 μm Si thickness.

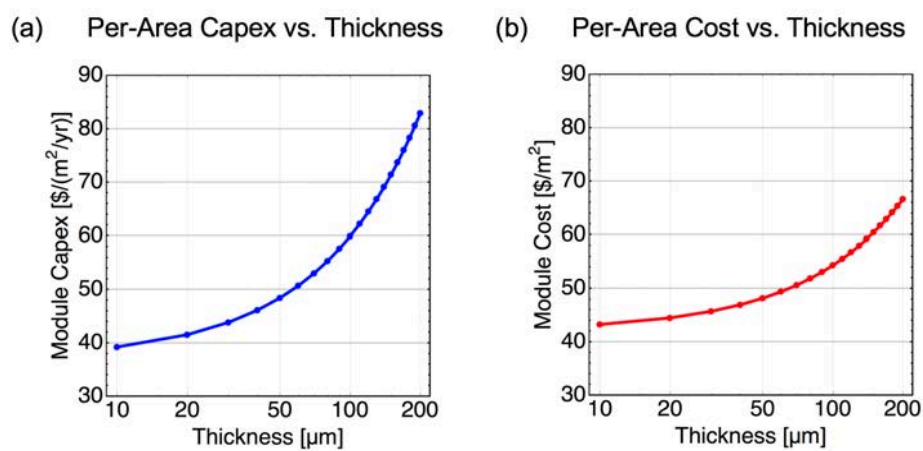

Figure S2: Per-area capex and cost versus silicon wafer thickness, which decoupled the effect of module efficiency.

## B. Device Simulation Parameters

The simulation parameters for *p*-type PV modules are shown in Table S1, and the *I-V* characteristics of the *p*-type solar cells in Figure S3. To compare with the *p*-type PV modules in Figure 2, we also simulated the *n*-type module efficiency in Figure S4.

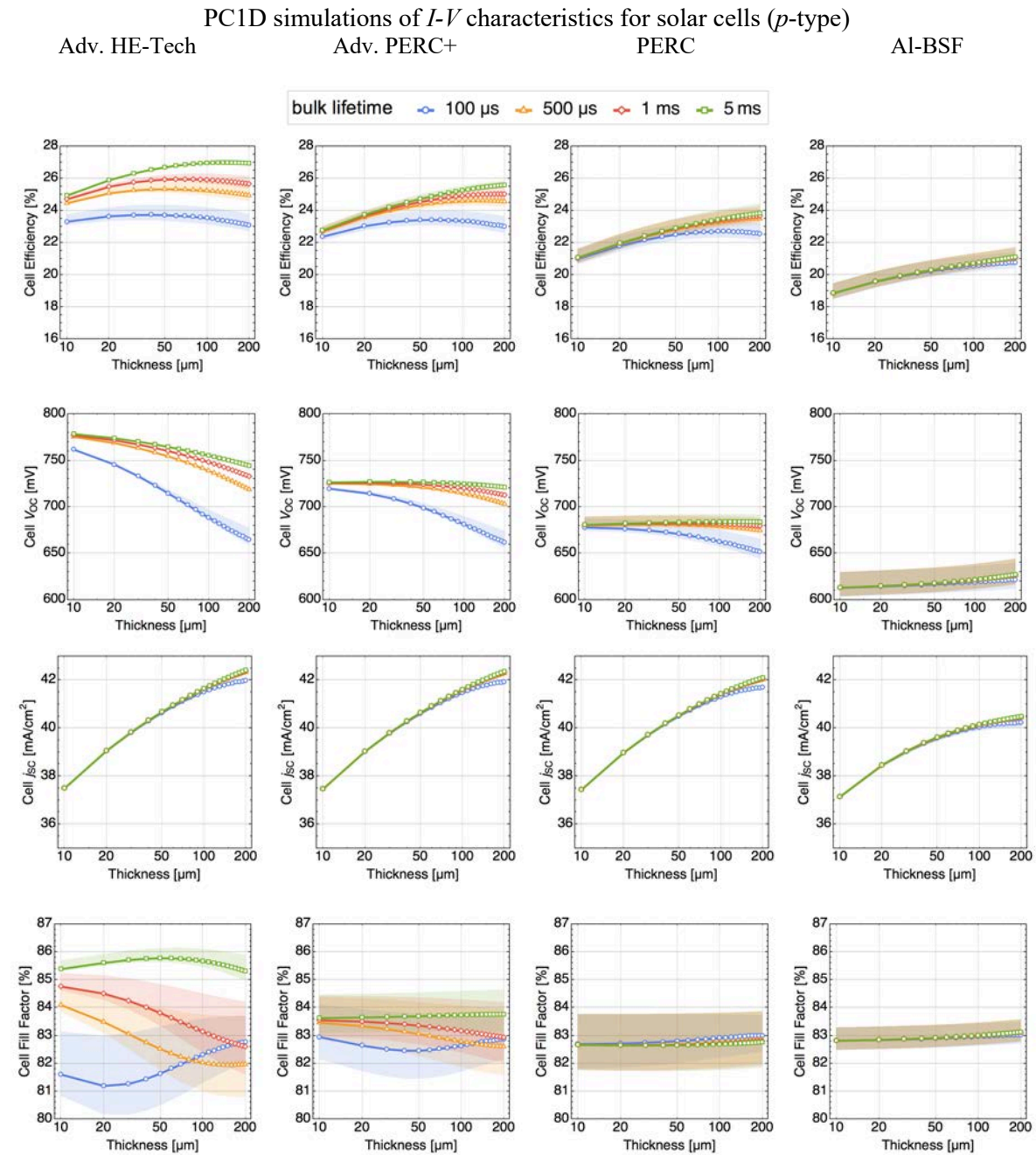

PC1D simulations of *I-V* characteristics for solar cells (*p*-type)

Figure S3: Solar cell efficiency, short-circuit current density $j_{SC}$, open-circuit voltage $V_{OC}$, and fill factor *FF* for solar cells with different technology concepts with *p*-type wafers.

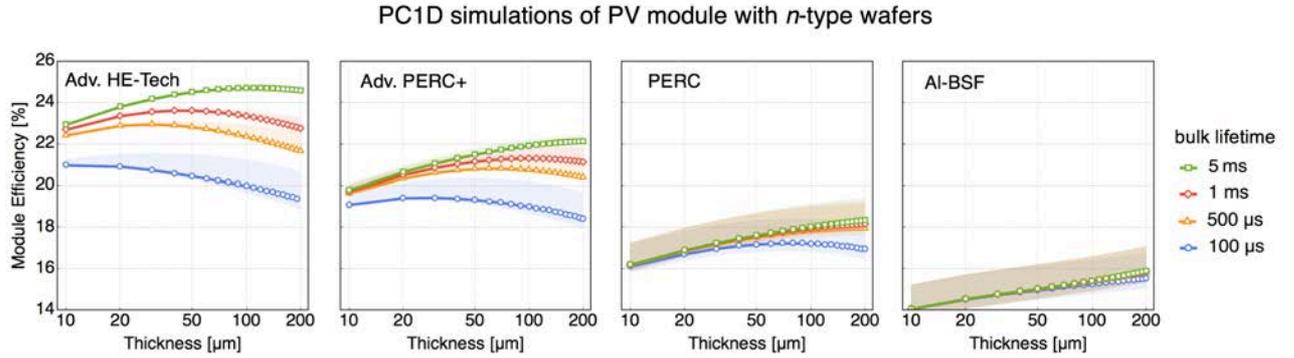

Figure S4: Simulated module efficiency with n-type wafers for different technological concepts and wafer bulk lifetimes. Base doping of *n*-type wafer is around 3 Ω·cm, with a typical range of 1 – 10 Ω·cm. The other parameters are set as the same as the p-type wafers in Table S1.

Table S1. Key parameters for PC1D device simulation of the four PV concepts in this study

| Simulation Parameters | Adv. HE-Tech | Adv. PERC+ | PERC | Al-BSF |
|---|---|---|---|---|
| Base resistivity, $\rho_{base}$ | 2 Ω·cm (*p*-type with a typical range of 1 – 3 Ω·cm) | | | |
| Emitter peak doping, $N_{em}$ | $6\times10^{18}$ cm$^{-3}$ | $6\times10^{19}$ cm$^{-3}$ | | |
| Front surface SRV, $S_{front}$ | 100 cm/s | 1000 cm/s | | |
| Rear surface SRV, $S_{rear}$ | 1 cm/s | 10 cm/s | 100 cm/s | 1000 cm/s |
| Rear surface reflectance, $R_b$ | 93% | | | 65% |
| Cell-to-module (CTM) efficiency factor, $f_{CTM}$ | 0.92 | 0.88 | 0.83 | 0.83 |
| Concept characteristics | • Ultra-low front and rear SRV<br>• Low Auger recombination<br>• Ultra-low CTM efficiency loss | • Very low rear SRV<br>• Very low CTM efficiency loss | • Low rear SRV<br>• Low CTM efficiency loss | • High rear SRV<br>• High parasitic absorption at the rear surface<br>• High CTM efficiency loss |

## C. Technoeconomic Analysis of Two Different Scenarios

The capex and cost of solar cell processing steps may increase for advanced concepts, due to the increase process complexity. We investigate a scenario where 1.5x capex and cost increase in cell processing of advanced PERC+ and 2.0x capex and cost increase in solar cell processing of advanced HE-Tech. Note that the capex of cell processing is 30% of the total current capex and the cost of celling processing is 19% of the total current cost (see Figure 1). In such a scenario shown in Figure S5a below, the cost curves for all the technological concepts are more or less overlap with each other, and the effect of the cost increase is thickness-invariant. However, the capex curves for advanced concepts shift above the baseline PERC scenario due to the relatively large proportion of capex required in the cell processing step. In addition, the decreasing thickness reduces the overall capex proportion of poly-Si and wafer capex, and therefore the capex gaps between advance concepts and PERC are dominated by the cell processing capex. As an consequence of cost increase, the LCOE curves shifts up as well as compared to Figure 3. However, even though all three cost curves are overlapped, we still observed LCOE reductions for advanced concepts due to the benefit of efficiency improvement.

Although we see a continuous decease of kerf loss in the past years, there is still a chance that the kerf loss starts to become thickness invariant (*e.g.*, due to technology limitation of wire sawing process).

In the second scenario shown in Figure S5b, we investigate constant 95 μm thick silicon kerf loss for all wafer thicknesses. Effectively, the thickness-invariant kerf loss results in less silicon saving per wafer for thinner wafer, and kerf loss starts to dominate the silicon usage for the thickness below 95 μm. We see the capex and cost curves get flattened (as compared to Figure 3) due to this effect, and the potential savings by thin wafers get much reduced. However, we still observe the optimum thicknesses for capex, cost and LCOE are located around 50 μm or less.

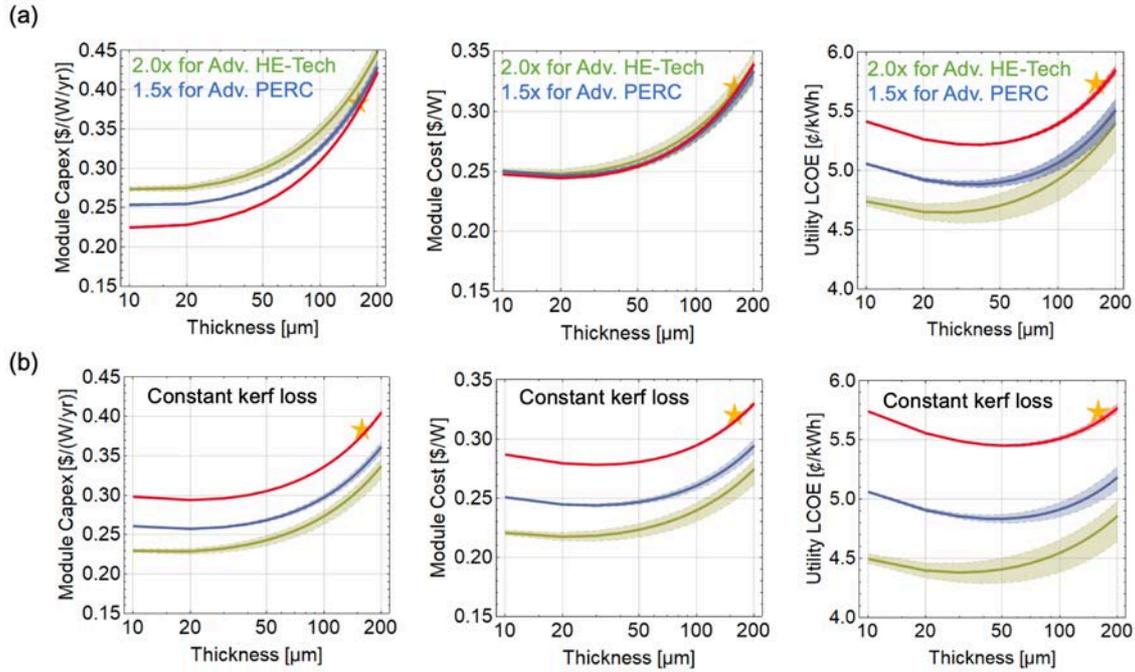

Figure S5: Cost analysis of two different scenarios: (a) capex and cost increase of cell processing of advanced concepts; (b) constant kerf loss of 95 μm thick silicon for all wafer thicknesses.

## D. Maximum Sustainable Growth over Years

Large increases in renewable energy generation are needed by 2030 to maximize the probability of maintaining global temperatures less than 1.5 – 2°C above pre-industrial levels. Due to the current and projected costs of PV, it is expected that PV will have a large role to play in producing this low-carbon energy, with deployment targets in multi-terawatt scale by 2030 [1]. In particular, the IPCC has stated that renewable energy generation must represent 20–30% of the total energy generation in 2030 to have a 25–75% chance of keeping atmospheric $CO_2$ between 430–480 ppm. If one third of this generation comes from PV with an average capacity factor of 20%, then based on International Energy Agency projections of global energy demand [2], the total of 7–10 TW PV needs to be deployed by 2030. To achieve this target, the manufacturing capacity must expand at a cumulative annual growth rate of 22–25%. The possibility of achieving the IPCC 2030 climate targets for the advanced concepts in thin silicon wafers was evaluated.

In Figure S6a, current baseline PERC with 160 μm wafer will not lead us to the 2030 goal in various scenarios. High debt ratio accelerates PV deployment in the short term and get us closer to the 7 TW by 2030, however the long-term growth for all the scenarios reaches a plateau around 7 TW due to demand constraint. In Figure S6b, the advanced HE-tech with 160 μm wafer (*i.e.*, no thickness reduction) has a faster growth but a similar trend for the baseline PERC. It also moves the plateau point above 10 TW. Because of very significant capex and cost reduction in advanced HE-Tech with 50 μm wafer, the cumulative PV deployment in Figure S6c maintains exponential growth well beyond 10 TW (the plateau point for this case is around 25 TW, and not shown in this figure).

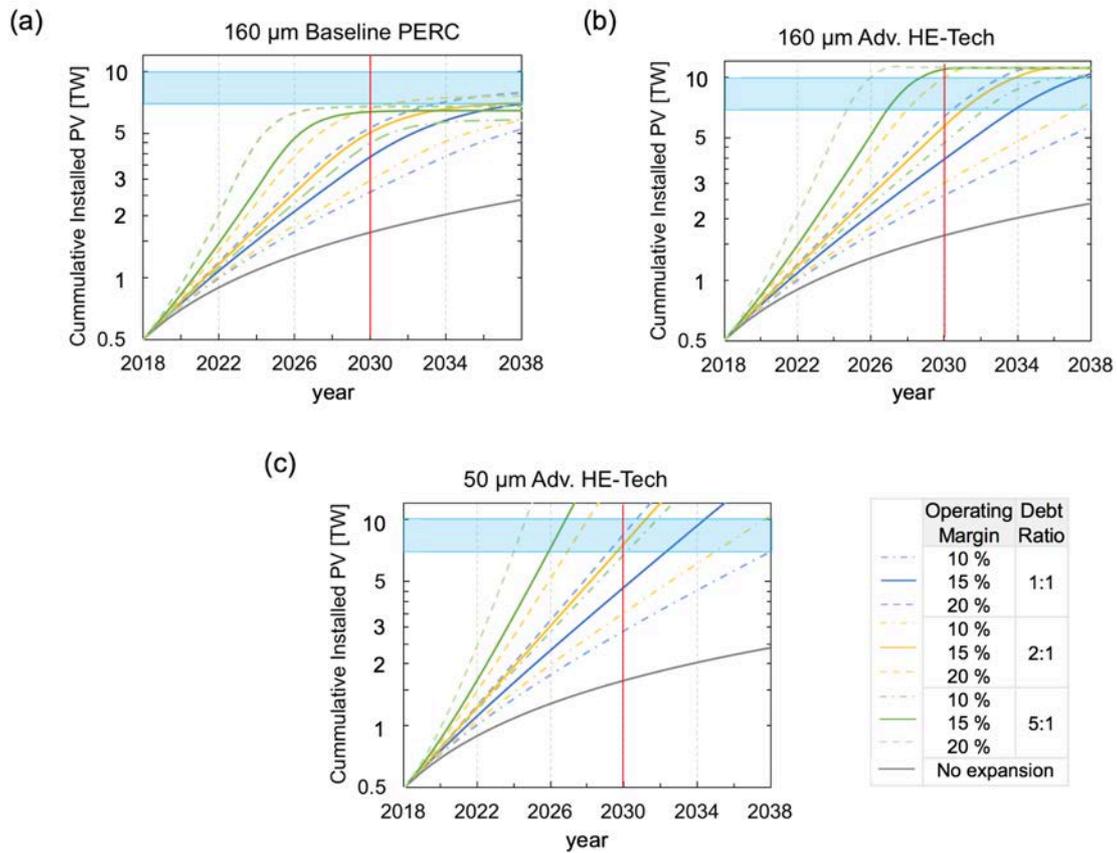

Figure S6: Cumulative PV installation over the next 20 years for adopting different technological concepts (using the growth model in Needleman et al. [3] with the updated cost values).

## E. Operating margins for PV companies

The operating margins over the past five years for eight selected PV companies were calculated from the annual financial data of each company published on Yahoo Finance [4] (see Figure S7).

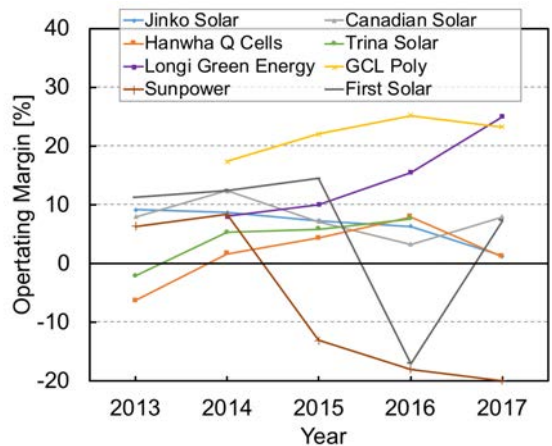

Figure S7: Operating margins of PV companies over the recent five years.

## F. Waterfall Breakdown of the Capex and Cost Difference

The waterfall breakdown of the current PERC with 160 μm wafer to the HE-Tech with 50 μm wafer. Under the assumptions in our model, we find in Figure S8 that the efficiency scales down both cost and capex universally, while the silicon savings has slightly larger impact on capex reduction (~27%) than cost reduction (~14%). As the results of uncertainty analysis, the error bars indicate the capex or cost change in response to ±5% change of every specific factor. The error bar for the advanced HE-Tech module indicates the overall uncertainty range of capex or cost if all the variables are changed by ±5%.

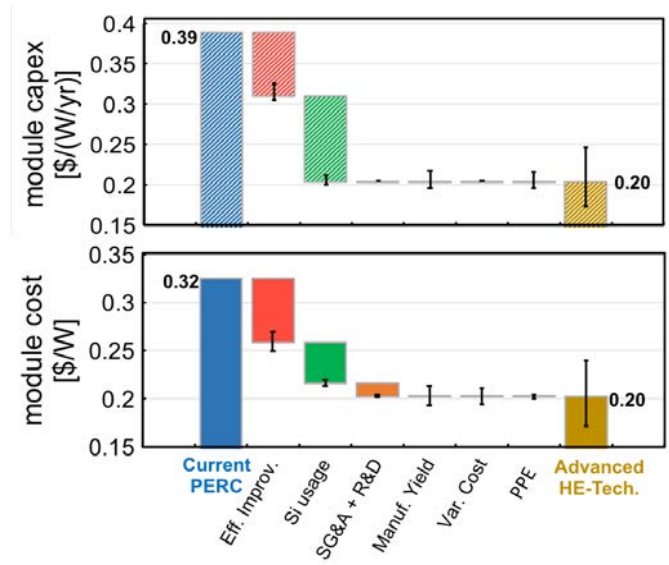

Figure S8: Waterfall chart of the capex and cost evolutions from the state-of-the-art PERC module with 19.0% efficiency and 160-μm-thickness wafer to the advanced concepts of advanced HE-Tech module with 23.8% efficiency and 50-μm-thickness wafer. The six factors in the breakdown are efficiency improvement, silicon saving, SG&A and R&D, manufacturing yield, direct variable cost and direct PPE. The bars with zero contributions indicate no change of the specific variable was considered in the cost model.

## G. Silicon Kerf Loss vs Wafer Thickness

The amount of silicon kerf loss is limited by the wire sawing process. In the past decade, kerf loss by wire sawing process as shown in Figure S9 is steadily reduced from 200 to 95 μm [5]–[7], with the projected additional reduction down to 60 μm (ITRPV predicted technology limit of wire sawing [8]). As mentioned briefly, achieving kerf loss below 60 μm may require novel technology of kerfless wafer growth, *e.g.*, epitaxial lift-off or direct wafer growth. Particularly, for wafers with 50 μm thickness and 28 μm kerf loss will necessarily require those kerfless technologies.

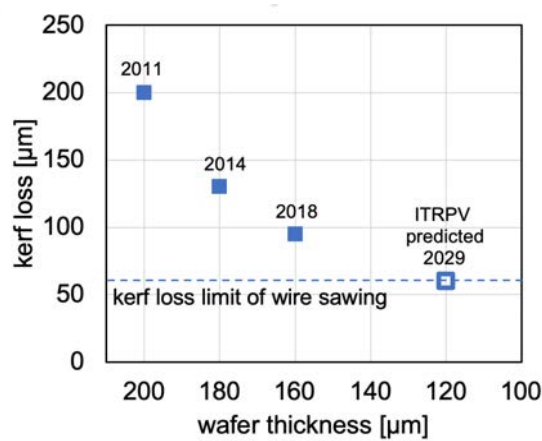

Figure S9: Silicon kerf loss vs wafer thickness for the past years from Ref. [5]–[7], and the ITRPV predicted kerf loss limit for wire sawing [8].

# Revisiting Thin Silicon for Photovoltaics: A Technoeconomic Perspective

Graphic Table of Contents

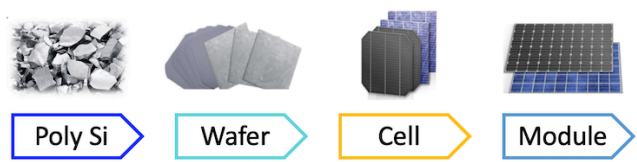

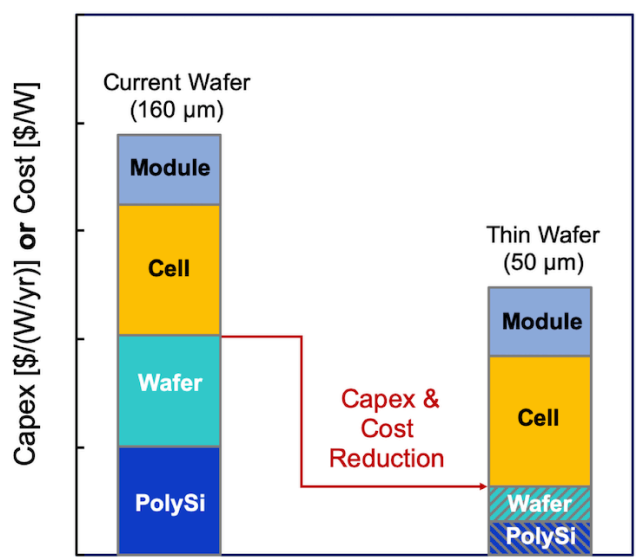

This technoeconomic analysis revisited the concept of thin silicon wafer for its potential cost benefits and technological challenges.